\documentclass[12pt,preprint]{aastex}

\def\msun{M_{\odot}}

\shorttitle{X-ray Spectra and Pulse Frequency Changes in SAX J2103.5+4545}
\shortauthors{Baykal, Stark, Swank }

\begin{document}


\title{X-ray Spectra and Pulse Frequency Changes in SAX J2103.5+4545 }


\author{A. Baykal }
\affil{Middle East Technical University, Physics Department, 06531 Ankara, Turkey}
\email{altan@astroa.physics.metu.edu.tr}

\author{M. J. Stark }
\affil{Department of Physics, Lafayette College, Easton, PA  18042,
USA}
\email{starkm@lafayette.edu}

\and

\author{J. H. Swank }
\affil{Laboratory for High Energy Astrophysics, NASA Goddard Space Flight Center,
Greenbelt, MD 20771, USA}
\email{swank@milkyway.gsfc.nasa.gov}



\begin{abstract}

The November 1999 outburst of the transient pulsar SAX~J2103.5+4545
was monitored with the large area detectors of the Rossi X-Ray Timing
Explorer until the pulsar faded after a year. The 358 s pulsar was spun up
for 150 days, at which point the flux dropped quickly by a factor of 
$\approx $ 7, the frequency saturated and, as the flux continued to
decline, a weak spin-down began. The pulses remained strong during the decay
and the spin-up/flux correlation can be
fit to the Ghosh \& Lamb derivations for the spin-up caused by
accretion from a thin, pressure-dominated disk, for a distance $\approx
3.2$ kpc and a surface magnetic field $\approx 1.2 \times 10^{13}$ Gauss. 
During 
the bright spin-up part of the outburst, the flux was subject to 
strong orbital modulation, peaking $\approx 3$ days after periastron of the eccentric 12.68 day orbit, 
while during 
the faint part, there was little orbital modulation. The X-ray spectra
were typical of accreting pulsars, describable by a cut-off power-law, with
an emission line near the 6.4 keV of K$_{\alpha}$ fluorescence from cool iron. 
The equivalent width of this emission did not share the orbital modulation, but
nearly doubled during the faint phase, despite little change in the column 
density. The outburst could have been caused by an episode of increased wind 
from a Be star, such that a small accretion disk is formed during each 
periastron passage. A change in the wind and disk structure apparently
occurred after 5 months such that the accretion rate was no longer modulated
or the diffusion time was longer. 
The distance estimate implies the X-ray luminosity  
observed was between $1 \times 10^{36}$ ergs s$^{-1}$ and 
$6 \times 10^{34}$ ergs s$^{-1}$, with a small but definite correlation
of the intrinsic power-law spectral index. 

\end{abstract}


\section{Introduction}

The transient X-ray source SAX J2103.5+4545
was discovered by the Wide Field Camera 
instrument on the BeppoSAX X-ray satellite 
during the outburst between February and September 1997  
 (Hulleman, in 't Zand, $\&$ Heise 1998).
The source showed 358.61 s  pulsations.
The X-ray spectrum was consistent with a power-law model. The   
 photon index was 1.27$\pm$0.14 and the absorption column density was 
3.1$\pm$1.4 $\times 10^{22}$ cm$^{-2}$.
 
Another outburst was detected 2 years later by the All-Sky Monitor (ASM) 
on the Rossi X-Ray Timing Explorer (RXTE). Pointed observations
were carried out with  RXTE. Doppler shifts of the pulsations
seen in these observations revealed that 
the orbital period is 12.68 days
(Baykal, Stark $\&$ Swank 2000a,b). 
The orbital parameters suggest that the source
has a high mass companion, but no suitable optical 
counterpart has been reported. Hulleman et al. (1998) pointed 
out a B star at the edge of BeppoSAX  error box, but its distance would imply
a luminosity too low to explain the spin-up that was seen in the initial
RXTE observations.  

SAX J2103.5+4545 continued to be active more than a year after 
the ASM detectors detected it in 1999 November. During the active interval 
the source was  monitored through regular pointed RXTE observations. 
In this work, we present results from analysis of the X-ray spectra as well
as new results from the pulse timing of the full set of observations. 
During this outburst, although the source is not a bright target
( $\le 30$ mCrab), 
it was possible to make very significant measurements
which have a bearing on several aspects of accreting pulsars.

\section{Observations and data analysis} 

SAX J2103.5+4545 was regularly observed between 
19/11/99 and 31/08/00.   
Approximately 3-4 daily observations took place per week 
with a total weekly exposure time of 10-15 ks. 
Between 12/09/00 and 02/12/00, the observations were separated 
by a few weeks.
In March 2001, the source began another outburst. RXTE carried out a 
few observations 30 days after. 
The total nominal exposure time of the observations was 521 ks.    
The results presented here are based on data collected with the
Proportional Counter Arrary (PCA) 
(Jahoda et al. 1996) and the High Energy X-ray Timing Experiment (HEXTE) 
(Rothschild et al. 1998). The PCA instrument consists of an array of five  
collimated xenon/methane multi-anode 
proportional  counters, although they were not all operated 
during some observations. 
The total effective area is approximately 6250 cm$^{2}$ and the field of view 
$\sim 1^{\circ}$  full width at half 
maximum (FWHM). The HEXTE instrument consists of two independent clusters of
detectors, each cluster containing four NaI(T1)/CsI(Na) phoswich scintillation
counters (one of the detectors in cluster 2 not used for spectral information)
sharing a common $\sim 1^{\circ}$ FWHM field of view.
The field of view of each cluster is switched
on and off source to provide background measurements. The net open
area of the seven detectors used for spectroscopy 
is 1400 cm$^2$. Each detector covers
the energy range 15-250 keV.

\subsection{X-ray light curves and spectra}

Background light curves and X-ray spectra 
were generated by using the background estimator models 
based on the rate of very large events (VLE), detector activation 
and cosmic X-ray emission. 
The background light curves were subtracted from the source 
light curve obtained 
from the binned GoodXenon  data.
Figure 1 presents the light curve per proportional counter unit (PCU). 
From this figure, 
it is clear that the count rate dropped after the first 150 days 
of observations. 
Background subtracted spectra for each active PCU were combined 
with FTOOLS software provided by the RXTE Guest Observer Facility of 
HEASARC. 
For the HEXTE data the background subtraction is straightforward 
since the HEXTE detectors are rocking on and off the source at 32 s intervals. 
Standard FTOOLS software 
was used for the HEXTE data reduction, including the deadtime correction.
We modelled the X-ray spectra of the PCA and HEXTE together
between 3 and 50 keV, using a power-law function  
with low-energy absorption (Morrison $\&$ McCammon 1983),
multiplied with an exponential high-energy cut-off function 
(White, Swank, $\&$ Holt 1983). 
The fits required an emission line consistent with the 6.4 keV of fluorescence
from cold iron. 
We checked that at the galactic longitude ($87.13^{\circ}$) and latitude
($-0.68^{\circ}$) of SAX J2103.5+4545 the contribution of galactic ridge 
emission is small. Pointed observations nearby and scans in the region
imply no more than 0.2 counts s$^{-1}$ per PCU, less than half of the
minimum rate detected at the end of the outburst, when the pulse period
was still measured, and too small to influence the spectrum 
extracted for the faint phase. 
We studied the 
X-ray spectra in the bright phase of the outburst ($<$ 150 days) and 
in the faint phase of the outburst ($>$ 150 days) independently.
Table 1 presents the best fit values.
The photon index and the line parameters changed significantly. 
The photon index  ($\Gamma$=1.27 $\pm$ 0.02) in the bright 
phase was consistent with the index found by BeppoSAX 
in the preceeding outburst. It was lower (flatter spectrum)
than the index in the faint phase ($\Gamma$=1.41 $\pm$ 0.04). 
The emission line 
energy did not change significantly, but the line flux halved, while the 
equivalent width increased. 
Figure 2 shows the joint X-ray spectra of RXTE/PCA and HEXTE
during the bright phase.
Thermal bremsstrahlung models did not give successful fits to the data.


In order to see the other variations in the spectral parameters on shorter 
time scales, we present in figure 3 the count rates  
and the best fit spectral parameters in the 3--20 keV range 
as a function of time, 
with the time resolution of 
approximately one week. We do not see significant changes in 
spectral parameters, with an exception 
of a decrease in power-law index at the brightest part of 
the observations, 100--150 days after the onset.

\subsection{Orbital Phase Spectra}

In our earlier study of this source,
we saw an increase in the count rate 
at  periastron passages (Baykal, Stark, $\&$ Swank 2000b).
This behaviour implied that the mass accretion rate 
onto the neutron star increases at  periastron 
passage due to the eccentric orbit (e=0.4 $\pm 0.2$). 
We now find that when we fold the
faint parts of the outburst on the orbital period 
of 12.68 days, 
we do not see any change in the count rate as a function of 
orbital phase. 
We confirmed this behavior by extracting the ASM data and folding at 
the orbital period. When we fold either the ASM or the 
PCA data in the bright phase, 
we see a significant orbital modulation, while for the faint phase we do not. 
Figure 4 
presents the best fit spectral parameters  
as a function of orbital phase in the bright phase of outburst. 
In this figure, an orbital phase of 0.4 $\pm$ 0.08  corresponds to the 
periastron passage of the neutron star, and clearly the 
X-ray flux is higher there, although it actually peaks 2-3 days later. 
The other spectral parameters do not show significant changes, with 
the exception of the power-law index. 
Figure 5 presents the power-law index, 
the equivalent width and the emission line flux as a function of X-ray flux.
In this figure, the power-law index decreases 
with increasing X-ray flux. 
In other words
the X-ray spectrum becomes harder when the X-ray flux is high
and softer when the X-ray flux is low.
The emission line flux is 
proportional to the overall X-ray flux, with the equivalent width consistent
with being constant.

\subsection{Pulse Frequencies and Pulse Frequency Derivatives}
  
 For the timing analysis, we corrected the light curves  to the
 barycenter of the Solar system 
 and then corrected for the binary motion of the pulsar 
 using the binary orbital
 parameters 
 deduced by Baykal, Stark, $\&$ Swank (2000b).
 We divided the total observation time span into 
 non-overlapping segments of 15--20 days.
 For each observation segment, we first obtained the nominal pulse frequency
 (or pulse period) by using a long Fourier transform (or power spectrum) 
 and then constructed 10-20 pulse profiles (one pulse profile for each RXTE orbit) 
 by folding the data at this nominal pulse period. Finally we
  found the pulse phase offsets (or pulse arrival times)
 by cross correlating the pulse profiles with a template 
chosen as
the most statistically significant
 pulse profile in each observation segment. We used the 
 harmonic representation of pulse profiles 
  (Deeter $\&$ Boynton 1985). In this technique, 
 the pulse profiles are expressed in terms of a harmonic series 
 and cross correlated with the template pulse profile. 
 The pulse phase offsets can be found in terms of a
 Taylor expansion,
\begin{equation}
\delta \phi = \phi_{o} + \delta \nu (t-t_{o})
+ \frac{1}{2} \dot \nu (t-t_{o})^{2}
\end{equation}
where $\delta \phi $ is the pulse phase offset deduced from the pulse
timing analysis, $t_{o}$ is the mid-time of the observations, $\phi_{o}$ is
the phase offset at t$_{o}$, $\delta \nu$ is the deviation from the mean
pulse frequency (or additive correction to the pulse frequency), and $\dot
\nu $ is the source's pulse frequency derivative.	
We fitted the phase offsets to the Taylor expansion 
given in equation 1. 
Figure 6 (middle panel) shows the resulting  15 pulse frequency 
derivatives as a function of the mid-times of the 
observations.
We made linear fits to the phase offsets 
(i.e. $\delta \phi = \phi_{o} + \delta \nu (t-t_{o})$) with  nearly 
weekly time resolution,
obtaining 31 pulse frequencies.  These 
are shown  in the upper panel in figure 6.  
The lower panel in figure 6 shows the X-ray fluxes associated with 
the pulse frequency derivatives. Figure 7 shows that the pulse frequency 
derivatives are clearly correlated with their associated X-ray flux values. 

\section{Discussion}

A correlation between spin-up rate and X-ray flux in different energy ranges 
has been observed in outbursts of 5 transient systems.
These systems are EXO 2030+375, 
(Parmar, White $\&$ Stella 1989, Parmar et al. 1989, 
Reynolds et al. 1996), 
2S 1417-62 (Finger, Wilson $\&$ Chakrabarty 1996), A 0535+26
(Bildsten et al. 1997, Finger, Wilson $\&$ Harmon 1996), 
GRO J1744-28 (Bildsten et al. 1997) 
and XTE J1543-568 
(In 't Zand, Corbet $\&$ Marshall 2001). 
All these sources were observed during  spin-up phases and 
the correlations between 
spin-up rate and X-ray luminosity were explained in terms of
accretion from an accretion disk. 
The outburst of SAX J2103.5+4545 started with 
a spin-up trend, made a transition to a steady spin rate  and 
then appeared to just begin a spin-down trend (see figure 7).
These measurements of 
SAX J2103.5+4545 are the first which have resolved the transition 
to spin-down.

If the accretion is from a Keplerian disk, 
at the inner disk edge the magnetosphere disrupts 
the Keplerian rotation of the disk, forcing matter to
 accrete along magnetic field lines. The inner disk 
 edge $r_{o}$ moves inward with increasing mass 
accretion rate. The dependence of the inner disk edge
$r_{o}$ on the mass accretion rate $\dot M$ 
may be approximately expressed as 
(Pringle $\&$ Rees 1972, Lamb, Pethick, $\&$ Pines 1973)
\begin{equation}
r_{o}= K \mu^{4/7}(GM)^{-1/7}\dot M^{-2/7}
\end{equation}
where 
$\mu$=$BR^{3}$ is the neutron star magnetic moment with  
$B$ the magnetic field and  $R$ the neutron star radius,
G is the gravitational constant, and $M$ is the 
mass of the neutron star. In this equation $K=0.91$ gives the Alfven 
radius for spherical accretion. 
 Then the torque estimate is given by 
 Ghosh \& Lamb (1979) as
 \begin{equation}
2\pi I \dot \nu = n(w_{s}) \dot M~l_{K},
\end{equation}
where $I$ is the moment of inertia of the neutron star,
$l_{K} = (GM)r_{o})^{1/2}$ is
 the specific angular momentum added by a Keplerian disk
 to the neutron star at the inner disk edge
 $r_{o} $;
\begin{equation}
n(w_{s}) \approx 1.4 (1-w_{s}/w_{c})/(1-w_{s})
\end{equation}
 is a dimensionless
 function that measures the variation of the accretion torque
as estimated by the fastness parameter
\begin{equation}
w_{s} =\nu /\nu _{K}(r_{o}) = (r_{o}/r_{co})^{3/2} =
 2 \pi 
K^{3/2} P^{-1}  (GM)^{-5/7}
    \mu ^{6/7} \dot M^{-3/7},
\end{equation}
where $r_{co}=(GM/(2\pi\nu)^{2})^{1/3}$ is the corotation radius 
at which the centrifugal forces balances the gravitational forces, 
$w_{c}$
is the critical fastness parameter at which the accretion
torque is expected to vanish. 
The critical fastness parameter $w_{c}$ has been estimated to be  
$\sim$ 0.35
and depends on the electrodynamics of the disk 
(Ghosh \& Lamb 1979, Wang 1987, Ghosh 1993, 
Torkelsson 1998).

The behavior of the dimensionless function $n$ as a function of 
$\omega_{s}$ can be understood as follows.
The accretion torque is the sum of the torque
produced by accretion of the 
angular momentum of the matter that falls onto the star (mechanical torque) 
and the torque contributed by the twisted magnetic field lines from 
the star that interact with the outer parts of the disk   
(magnetic torque). The mechanical torque 
always acts to spin-up a star rotating in the same sense 
as the disk flow, whereas torque from the magnetic stresses can have either
 sign, since the azimuthal pitch of 
the stellar magnetic field lines that interact with Keplerian flow in the 
disk changes sign at the corotation radius $r_{co}$. 
The torque 
from the magnetic field lines threading the disk between 
the inner disk edge $r_{o}$ and 
corotation radius
$r_{co}$ is positive, whereas the 
contribution of torque from the magnetic field lines threading the disk outside 
the corotation radius $r_{co}$ is negative.
 Therefore, either 
spin-up or spin-down torque is possible in this model as a net effect of
the balance between the two contributions. 
Equation 4 gives an analytic expression approximating numerical 
calculations of the dimensionless torque (Lamb 1988, Ghosh 1993).
The torque will cause a spin-up
if the neutron star is rotating slowly
 ($w_{s}~<~w_{c}$) in the same
sense as the circulation in the disk.
 Even if the
neutron star is rotating in the same sense as the disk flow,
the torque will be in the direction of spin-down if the neutron 
star is rotating too rapidly
 ($w_{s}~>~w_{c}$).

The accreted material  
will produce X-ray luminosity at the neutron star surface at the rate
\begin{equation}
L =  GM \dot M /R
\end{equation} 
From 
equations 2,3 and 6, the rate of spin-up is related to the X-ray 
intensity through 
\begin{equation}
\dot \nu \propto n(w_{s}) L^{6/7} =  n(w_{s}) (4 \pi d^{2} F)^{6/7},
\end{equation}
where $d$ is the distance to the source and $F$ is the X-ray flux. 

Since the source distance and magnetic field
are not known, 
we fit the Ghosh \& Lamb model of the pulse frequency derivative
for these two parameters. Because the 
power index between the spin-up rate and  the X-ray luminosity 
is a result of the theoretical model, as a test of the model we also
fit for this index, obtaining 
0.75 $\pm$ 0.13, which is consistent with the value 6/7 expected in the 
model.
We obtain for the distance to the source  3.2 $\pm$ 0.8 kpc and for
the magnetic field 
(12 $\pm$ 3) $\times 10^{12}$ Gauss.
As seen in figure 6, the points corresponding to spin-up and spin-down 
are consistent with falling on the same curve.
With the estimated distance of 3.2 $\pm$ 0.8 the source could be in a  
star formation 
region in the Perseus arm at approximately  4 kpc 
(Vogt \& Moffat 1975, Georgelin \& Georgelin 1976.
The X-ray flux at the peak of the outburst
 $\sim 6.4 \times 10^{-10}$ erg s$^{-1}$ cm$^{-2}$
corresponds to an accretion luminosity of  
$\sim 8.8 \times 10^{35}$ erg s$^{-1}$ and 
a mass accretion rate $ \sim 6.5 \times 10^{15}$ g s$^{-1}$).
The magnetic field estimate $\sim 12 \times 10 ^{12}$ Gauss
 corresponds to a cyclotron
emission line at $\sim$ 140 keV, which is too high to be detected 
by HEXTE, because of the low count rate of the source 
at this energy range.

 During the decrease of mass accretion rate $\dot M$, 
 the fastness parameter increases as $w_{s} \sim \dot M^{-3/7}$.
 When the fastness parameter reaches
 $w_{s}=(r_{o}/r_{co})^{3/2} \sim $ 0.35, the inner edge of the disk 
 will be still be well inside the corotation radius $r_{o} \sim 0.5 r_{co}$.
 In this case, even if the net torque on the neutron star vanishes,
 a large pulse fraction change is not expected, since the polar 
 cap region at which the material accretes and radiates
 will not be changed significantly. 
 Figure 8, presents the 3-20 keV pulse profiles at
 the bright and faint phases of the outburst. It is clearly
 seen that there is no significant changes in the pulse
 profiles.

 Further decrease of the mass accretion rate eventually will expand the 
 magnetospheric radius $r_{m} \approx  r_{o}$ 
 to the corotation radius  
 $r_{m} \sim r_{co}$.
 Then, some of the material will be accelerated
 to super-Keplerian velocities and can not easily be accreted. 
 It may be expelled from the system.
 Accretion of material will carry angular momentum and tend to spin up 
 the neutron star, while the expulsion of matter
 will extract angular momentum from 
 the star. These forces tend to bring the neutron star 
 into rotation  at the equilibrium period.
 It is expected that accretion is eventually centrifugally inhibited.
 In this propeller regime the neutron star would rapidly spin-down
 and the X-ray luminosity might be produced by 
 the release of gravitational energy at the magnetosphere 
 (King $\&$ Cominsky 1994, Campana et al. 1995). 
 X-ray luminosity from the magnetospheric emission would be reduced 
 by a factor of $10^{3}-10^{4}$ (Corbet 1996). 

We have assumed here that the accretion is still through a disk. Given 
that the luminosity is very low 
($< 10^{36}$ ergs s$^{-1}$) and
the companion probably an early type star, 
it is not obvious that the accreting material
would be able to form a disk, that is, have enough
angular momentum to circularize outside the magnetosphere. Wang (1981)
showed that this would be the case if the wind velocity is relatively
slow ($< 500$ km s$^{-1}$) rather than fast ($> 1000$ km s$^{-1}$).
The companion is likely to be a Be star, although not yet identified,
because these have episodes of
dense slow winds in the equatorial plane (Waters at al. 1988; 
Li $\&$ van den Heuvel 1996). 

However, if the wind is slow enough that the velocity relative to the
neutron star is dominated by the orbital velocity, the accretion would
be relatively independent of phase or
tend to peak at apastron rather than near periastron as
observed. Diffusion through a disk might shift the time of peak X-ray
luminosity, but if the shift is as long as a week, the diffusion would
also reduce the amplitude of modulation. It is a fast wind with a
spherical outflow for which the rate of capture would peak at
periastron.  What is observed appears to be the opposite of a simple
picture of an episode of equatorial slow wind decaying to a fast wind.
The phasing of the modulation shown by SAX J2103.5+4545 has
been seen for other sources, such as V0332+53 (Waters et
al. 1989), in one outburst in which it appeared consistent with prompt
response to a wind of about 300 km s$^{-1}$.

The simple picture does not capture the complexity of  
Be Star winds. The influence of a
wind rotational velocity and a range of velocity laws (or density
dependence on distance from the star), in addition to the wind
velocity, were studied by Waters et al. (1989).
For the SAX~J2103.5+4545 outburst that we monitored,   
a wind velocity of 200-300 km s$^{-1}$ 
and a typical mass loss rate of 10$^{-8} \msun$ yr$^{-1}$
of a Be star could give the implied peak X-ray luminosities, the 
orbital modulation, and the phase dependence. But the additional 
parameters would be needed and several different regimes seem
possible.

The drop in accretion rate which we observed 
in SAX~J2103.5+4545
coincided with change in
the rate of angular momentum exchange. Presumably, both are caused by
changes in the wind structure. Exactly what these are is not obvious. 
The smooth behavior of the angular acceleration of the pulsar, as
opposed to erratic (random walk) variations like those of Vela X-1
(Bildsten 1997), for example, implies the accretion is still via a
disk.  The accretion rate has not yet fallen so low that the propellor
effect comes into play.  The disk accretion should moderate orbital
dependence. For there to be a disk and yet strong orbital modulation
at the peak of the outburst, the disk must be small. The lack of
modulation at the end of the outburst would be consistent with a disk
that is more extensive, through which the diffusion is longer.

Li and van den Heuvel (1996) considered the spin period versus binary
period in the diagram first constructed by Corbet (1984). The accretion
from a slow wind to a neutron star spinning in equilibrium 
appears responsible for the main correlation of the Be
star systems. Accretion from a fast wind is manifested by
supergiants with longer pulse periods for a given orbital 
period and for Be stars in certain phases. 
The values of spin and orbital period place SAX~J2103.5+4545 
among these sources, but a supergiant is usually a steady rather
than a transient source. SAX~J2103.5+4545 can be viewed as not having yet
achieved equilibrium. The transient episode decreased the pulse period by 
$\approx 0.9$ s. If this occurs every 1.4 yr, the source would spin up to 
the equilibrium line (1--2 s for a 13 d orbital period) in $< 600$ yr.
The frequency value obtained in the 2001 outburst was $1.4 \times 10^{-3}$ mHz
{\em above} the last previous measurement. It is consistent with an average
spin-up rate during the first 30 days (unobserved) of the ourburst being an 
average of $5 \times 10^{-13}$ Hz s$^{-1}$, about the same as in the beginning
of the 1999 outburst. The spin-down between the last monitored value in 2000
and the beginning of the 2001 outburst would have been only 
about $10^{-4}$ mHz at the rate observed. The spin rate apparently ratchets 
up with each outburst.

The X-ray spectra in principal can carry information about the 
disk and the wind as well as about the flow onto the neutron star. 
The 6.4 keV line, which comes from fluorescence from relatively cold Fe, could 
come from all of these sources. Fluorescent Fe K$_{\alpha}$ is common to many
high-magnetic-field  binary pulsars. The equivalent widths in the bright and 
faint phase spectra of SAX J2103.5+4545 are similar in magnitude to those 
observed in spectra of Vela~X--1 and GX~301--2 (summarized in Nagase 1989),
the accretion rate in SAX J2103.5+4545 is 10--100 
times lower and there is no varying column density indicating formation of 
a shell around the source. Yet for Vela X-1, where 
the neutron star is eclipsed,  the behavior of the line as a function
of binary phase implied a source close to the neutron star in addition to 
the wind. In the case of SAX~J2103.5+4545, the lack of orbital dependence
of the equivalent width during the bright part of the outburst 
again suggests the line is produced close to the neutron star 
where the geometry of cool material can be 
independent of the orbital position. In the decaying part of the outburst,
the equivalent width is nearly twice as large. The moderate column density
implies the material is seen in reflection rather than transmission. 
The cool fluorescing material may subtend a  larger solid angle, or a 
second component is excited for the new type of wind. The data could not 
significantly limit  orbital modulation of the line emission 
during this part of the outburst.

The spectral index of the intrinsic X-ray spectrum appears to be
consistently correlated with the flux, harder for higher flux, both in
the variation with orbital phase and the variation during the
outburst. The fact that the pulse amplitude and shape did not change
suggests that the X-rays continue to be produced in flow to the
neutron star and in that case the X-ray flux seems likely to be
proportional to the rate of accretion onto the surface.  Intrinsic
X-ray spectra for low luminosity sources $\dot M < 10^{17}$ g s$^{-1}$
have been modelled (Meszaros et al. (1983); Harding et al. (1984)).
The continuum spectrum has not appeared to be very sensitive to the
mass accretion rate, but the contributions of the cyclotron line
emission were not included and could make a difference if Compton
scattering of line photons affects the continuum below the line
energy. There has appeared to be at least a relation between the
cut-offs observed in pulsar spectra and the cyclotron resonance
energies (Makashima \& Mihara 1992; see however 
Mihara, Makashima, \& Nagase 1998). 
The high magnetic field obtained from the spin rate
dependence puts the cyclotron resonance energy farther above the
cut-off energy than is the case for pulsars with identified resonance
features and would give SAX~J2103.5+4545 the highest field of the accreting
pulsars (Orlandini \& Dal Fiume 2001). 
The magnetic field estimate depends on the critical fastness
and the balance between the magnetosphere and the accretion flow.
Quantitative estimates seem likely to be subject to dependence 
of the model on assumptions about the nature of the disk, for example.
Understanding the X-ray spectra and flux
correlations appears to require better models of the spectral
formation and also the magnetospheric physics. 

RXTE's monitoring of this transient pulsar allowed the 
spin rate to be tracked while the luminosity declined to 
$6 \times 10^{34}$ ergs s$^{-1}$. The spin-rate responded smoothly to the 
flux, spinning up, coming to equilibrium and then reversing sign, just as
the Ghosh \& Lamb characterization of the magnetospheric interaction 
predicted. While the magnetic field and distance that come out of 
fitting the data may be estimates subject to model 
dependence of the spin-up/flux relation, this source demonstrates 
the possibility of determining system
parameters from such observations. The system is also interesting in that 
it appears to be  different from better known ones and provides 
information about uncharted parts of parameter space of pulsars in binaries. 
Detailed study of the orbital phase dependence during the outbursts 
of this source will provide constraints on the stellar orbit flows 
from which this neutron star accretes.

\clearpage


\figcaption{Count rate per PCU history of SAX J2103.5+4545 in 3-20 keV 
\label{fig1}}

\figcaption{The RXTE/PCA-HEXTE spectra of SAX J2103.5+4545. The lower 
panel shows the residuals of the fit in terms of 
$\chi^{2}$ values.
\label{fig2}}

\figcaption{Top panel: Count rate per PCU history of SAX J2103.5+4545 in 3-20 keV. 
The errors (not indicated) are typically 1$\%$. Second through seventh panel:
time history of hydrogen column density, centroid energy of the emission line, 
equivalent width of the emission line, power-law photon index, cut off energy and 
e-folding of cut off energy. For some spectra the emission line or
cut off energy measurements are not significant due to  
small numbers of counts and these data points are not included in these panels.
 Bottom panel: reduced $\chi ^{2}$ statistic of 
the relevant spectral fits.   
 \label{fig3}}

\figcaption{Top panel: X-ray flux of SAX J2103.5+4545 in 3-20 keV as a function of orbital phase.
 Second through seventh panel:
hydrogen column density, centroid energy of the emission line,
equivalent width of the emission line, power-law photon index,
 cut off energy and e-folding of cut off energy
 as a function of orbital phase.
 Bottom panel: reduced $\chi ^{2}$ statistic of
the relevant spectral fits.
 \label{fig4}}

\figcaption{First through third panel: 
Power-law photon index, equivalent width of the emission line and 
emission line flux as a function of X-ray flux.
 \label{fig5}}

\figcaption{First through third panel: Pulse frequency, pulse frequency derivative 
and X-ray flux history  
 of SAX J2103.5+4545 as a function of time. The increase at the end 
is 30 days after the start of the next outburst.
 \label{fig6}}

\figcaption{Pulse frequency derivative versus X-ray flux 
of SAX J2103.5+4545, 
solid line presents the best fitted model to the data.
 \label{fig7}}

\figcaption{Normalized pulse profiles for
 the bright (dashed one) and 
faint phases (solid one) of the outburst in 3-20 keV. 
 \label{fig8}}

\clearpage

\begin{figure}
\plotone{f1.ps}
\end{figure}

\clearpage

\begin{figure}
\plotone{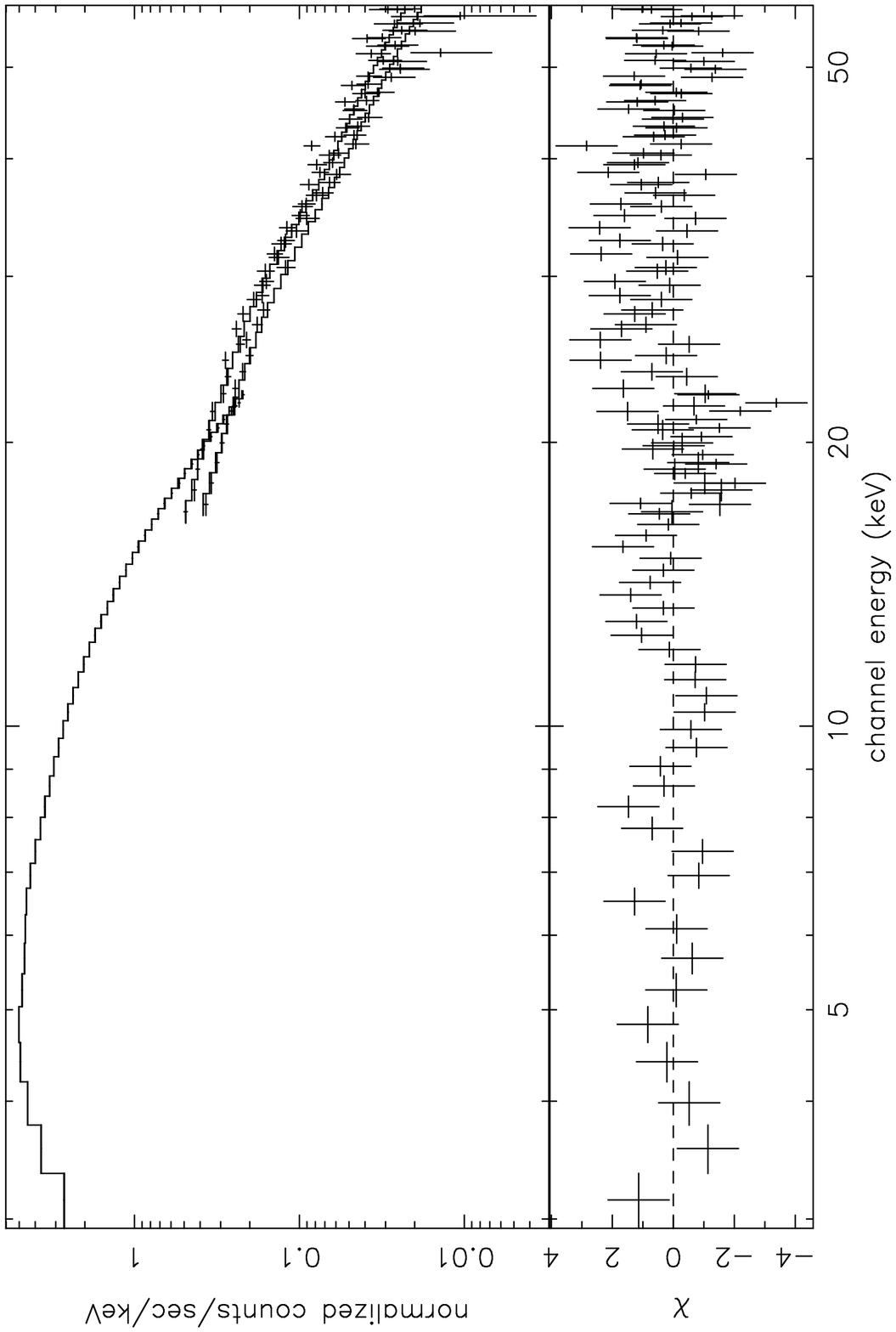}
\end{figure}

\clearpage

\begin{figure}
\plotone{f3.ps}
\end{figure}

\clearpage 

\begin{figure}
\plotone{f4.ps}
\end{figure}

\clearpage

\begin{figure}
\plotone{f5.ps}
\end{figure}

\clearpage

\begin{figure}
\plotone{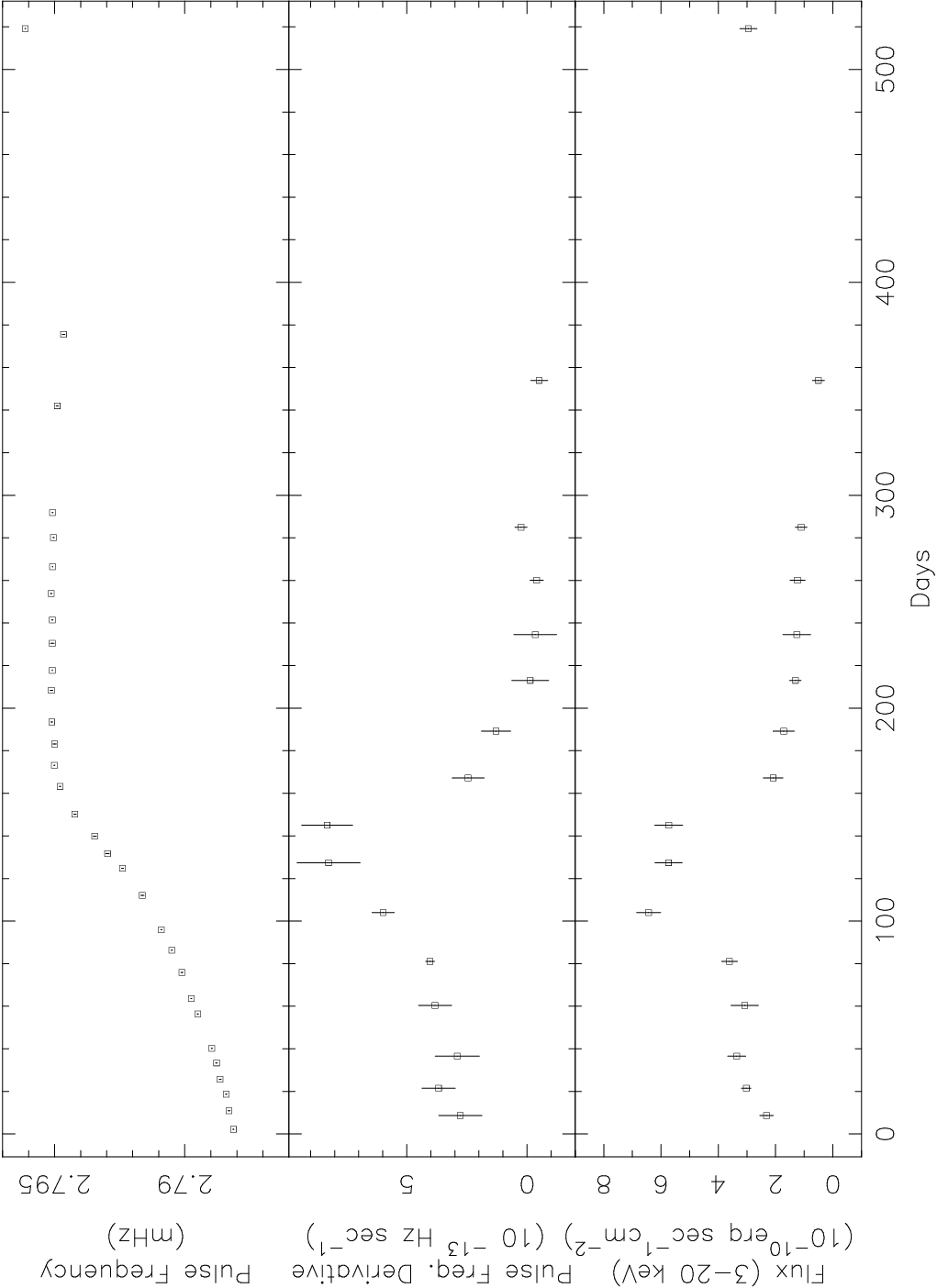}
\end{figure}

\clearpage

\begin{figure}
\plotone{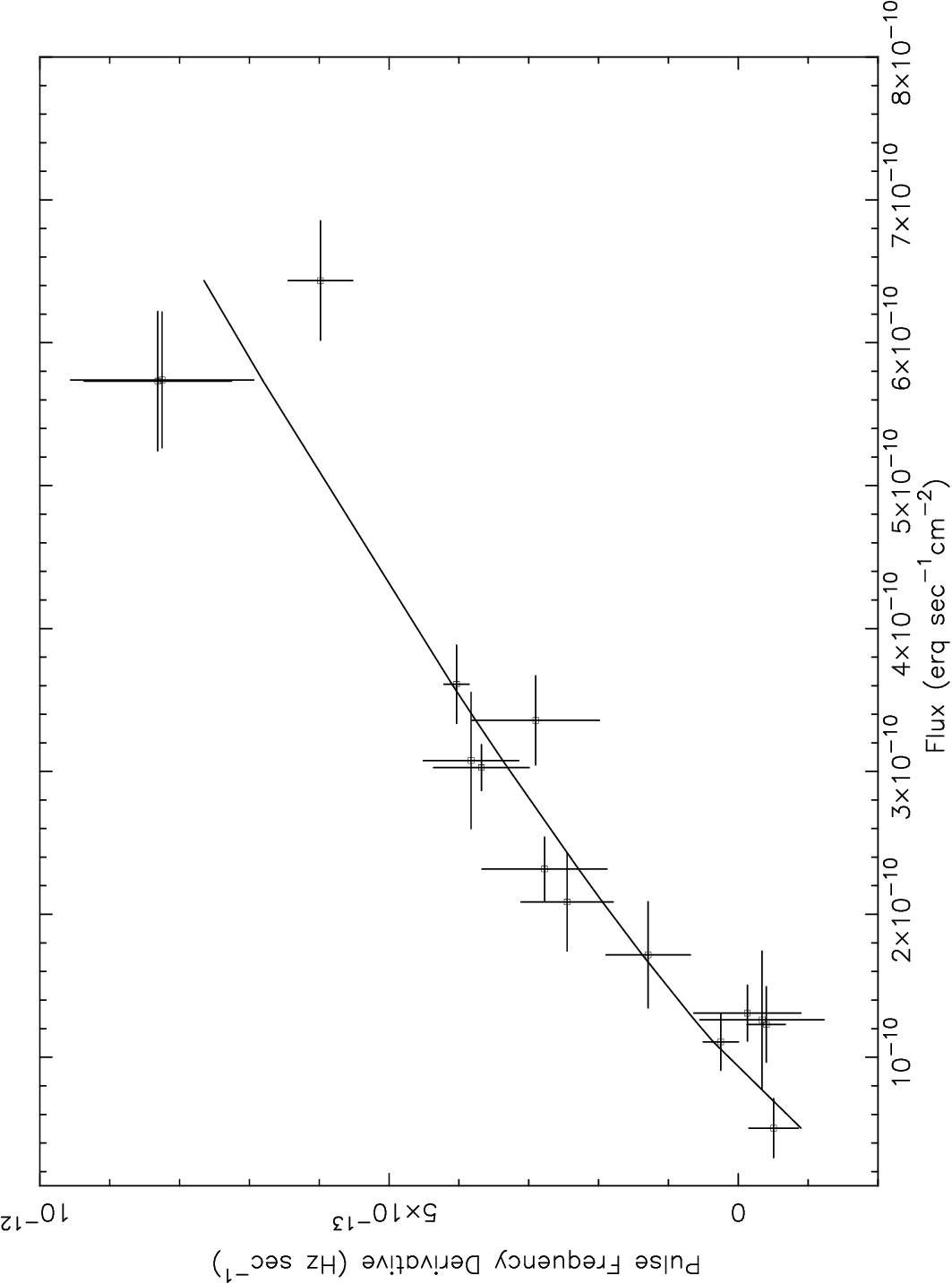}
\end{figure}


\clearpage

\begin{figure}
\plotone{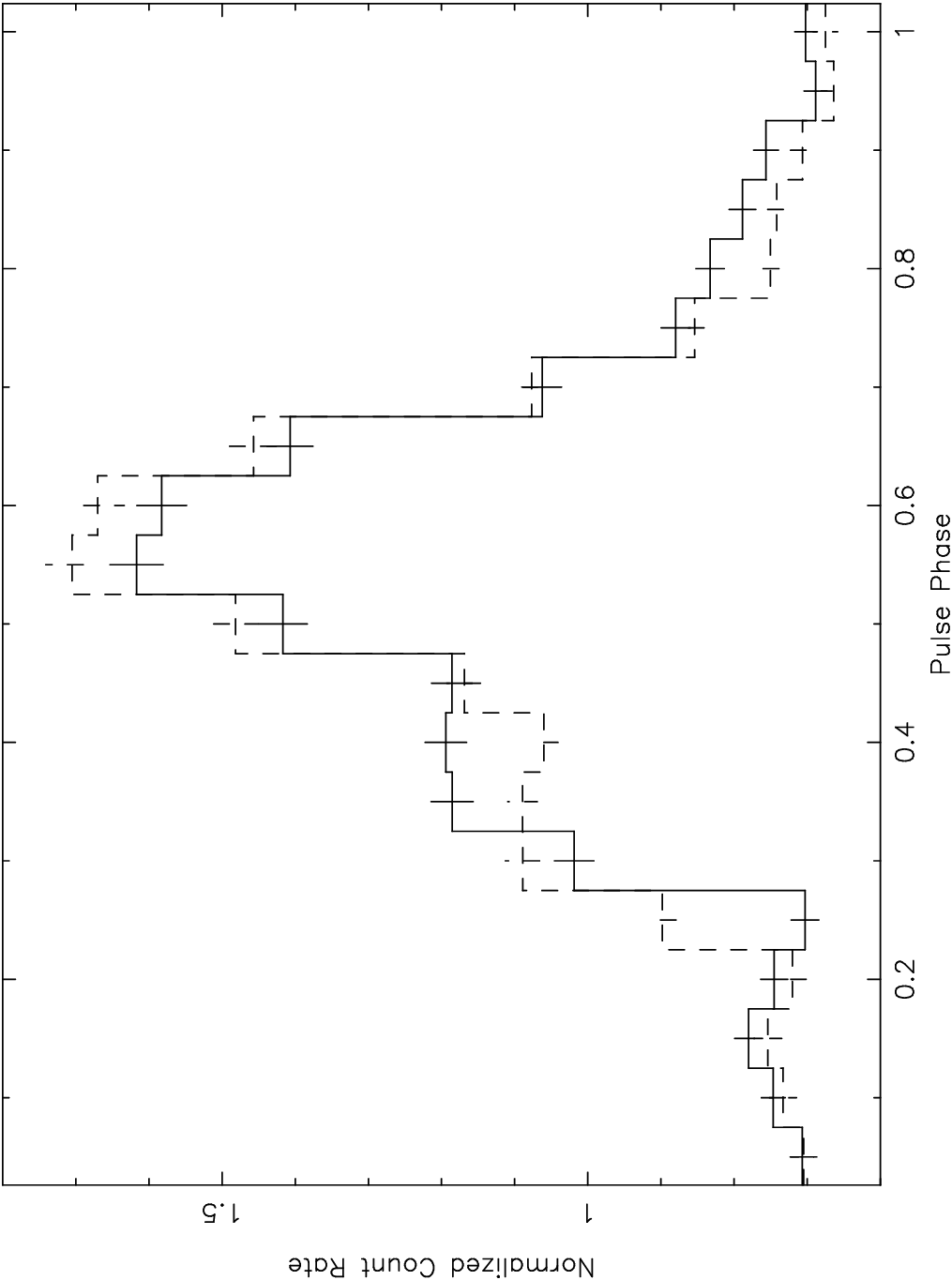}
\end{figure}






\clearpage

\begin{table}
\caption{Spectral Fit Parameters for RXTE observations$^{a}$}
\label{Pri}
\[
\begin{tabular}{c|c|c}   \hline
                        & Bright state & Faint state \\ \hline 
N$_{H}$ (10$^{22}$cm$^{-2}$) & 3.80 $\pm$ 0.10 & 3.65 $\pm$ 0.17 \\
Fe line center energy (keV) & 6.43 $\pm$ 0.02 & 6.44 $\pm$ 0.06 \\
Fe line width (eV) & 165 $\pm$ 10  & 306 $\pm$ 57 \\
Fe line intensity photons cm$^{-2}$ sec $^{-1}$ & (7.61 $\pm$ 1.35 )$\times$
 $10^{-4}$  & (3.12 $\pm$ 0.58)$\times$10$^{-4}$\\
Photon Index & 1.27 $\pm$ 0.02 & 1.41$\pm$0.04 \\
Cutoff Energy (keV) & 7.89 $\pm$ 0.28 & 7.89$^{b}$ \\
Folding Energy (keV) & 27.10 $\pm$ 0.94 & 27.1$^{b}$ \\
Power-law Norm. cm$^{-2}$ at 1keV & (4.84 $\pm$
0.16)$\times$ $10^{-2}$	& (1.41 $\pm$ 0.05) $\times$ $10^{-2}$ \\
Reduced $\chi^2$ & 1.17 (d.o.f. = 124) &  1.2 (d.o.f = 40) \\ \hline
\end{tabular}
\]
$^{a}$ Uncertainties in the spectral fit
parameters denote 68$\%$ confidence.\\
$^{b}$ HEXTE data are not used at faint state fit because of 
low count rates. In the fitting procedure, cutoff energy and folding energy
 are fixed according to values found in bright state.
\end{table}

\end{document}